\documentclass[10pt,onecolumn]{article} 

\usepackage{amssymb}
\usepackage[cmex10]{amsmath}
\usepackage{paralist} 
\usepackage{graphicx}
\usepackage[tight,footnotesize]{subfigure}
\usepackage{algpseudocode}
\usepackage[font=footnotesize]{caption} 
\usepackage{fullpage}

\renewcommand\footnotemark{}

\begin{document}

\title{Analysis Based Blind Compressive Sensing}

\author{Julian~W\"ormann,
        Simon~Hawe,
        and~Martin~Kleinsteuber
\thanks{Copyright \copyright~2013 IEEE. Personal use of this material is permitted. However, permission to use this material for any other purposes must be obtained from the IEEE by sending a request to pubs-permissions@ieee.org.}%
\thanks{All authors are with the Department
of Electrical Engineering and Information Technology, Technische Universit\"at M\"unchen, 80290 M\"unchen Germany, e-mail: julian.woermann@mytum.de, simon.hawe@tum.de, kleinsteuber@tum.de.}
\thanks{This work has been supported by the German Federal Ministry of Economics and Technology (BMWi) through the project KF3057001TL2 and by the Cluster of Excellence CoTeSys - Cognition for Technical Systems, funded by the German Research Foundation (DFG).}%
\thanks{Digital Object Identifier 10.1109/LSP.2013.2252900}}


%


\date{}
\maketitle

\begin{abstract}
In this work we address the problem of blindly reconstructing compressively sensed signals by exploiting the co-sparse analysis model. 
In the analysis model it is assumed that a signal multiplied by an analysis operator results in a sparse vector.
We propose an algorithm that learns the operator adaptively during the reconstruction process.
The arising optimization problem is tackled via a geometric conjugate gradient approach. 
Different types of sampling noise are handled by simply exchanging the data fidelity term. 
Numerical experiments are performed for measurements corrupted with Gaussian as well as impulsive noise to show the effectiveness of our method. 
\end{abstract}


\begin{center}
\small{\textbf{Index Terms}}

\vspace{.5em}
\small{Analysis Operator Learning, Blind Compressive Sensing, Optimization on Matrix Manifolds.}
\end{center}

%
\vspace{2em}

\section{Introduction}
\label{sec:1}
\subsection{Regularization in Compressive Sensing}
\label{subsec:11}
In recent years, Compressive Sensing (CS) has influenced many fields in signal processing. Basically, the theory states that if an unknown signal ${\bf{s}} \in \mathbb{R}^{n}$ can be sparsely represented, only a few $m < n$ linear and non-adaptive measurements ${\bf{y}}\in\mathbb{R}^{m}$ of the signal suffice to accurately reconstruct it. Denoting the measurement vectors by $\{ \phi_{i} \in\mathbb{R}^{n} \}_{i = 1}^{m}$, the measurement process can be compactly written as   
\begin{equation}
\label{eq:measurements}
  {\bf{y}} = [ \phi_{i}, \ldots, \phi_{m} ]^{\top} \bf{s} + \bf{z} = \bf{\Phi} \bf{s} + \bf{z},
\end{equation}
where ${\bf{\Phi}} \in \mathbb{R}^{m \times n}$ is the measurement matrix, and  ${\bf{z}}\in\mathbb{R}^{m}$ constitutes possible sampling errors. Due to the reduced dimensionality, reconstructing $\bf{s}$ from the measurements is ill-posed in general, and cannot be done by simply inverting  ${\bf{\Phi}}$. 
However, additional model assumptions on $\bf{s}$ may help to find a solution. 
In this context, the \emph{sparse synthesis-approach} and the \emph{co-sparse analysis-approach} \cite{elad-invprob-07} have proven extremely useful.
In the sparse synthesis approach it is assumed that a signal can be decomposed into a linear combination of only a few columns, called atoms, of a known dictionary ${\bf{\mathcal{D}}}\in\mathbb{R}^{n \times d}$  with $d \geq n$, i.e.  $\bf{s} = {\bf{\mathcal{D}}} \bf{x}$ with ${\bf{x}}\in\mathbb{R}^{d}$ being the sparse coefficient vector. 
Many algorithms for solving the synthesis problem exist, cf. \cite{tropp-poti-2010} for an extensive overview.

The co-sparse analysis approach is a similar looking but yet very different alternative to tackle the CS problem. Its underlying assumption is that a signal multiplied by an \textit{analysis operator} ${\bf\Omega} \in \mathbb{R}^{k\times n}$ with $k \geq n$ results in a sparse vector ${\bf\Omega s} \in \mathbb{R}^k$. If $g\colon \mathbb{R}^{k} \to \mathbb{R}$ denotes a function that measures sparsity, the analysis model assumption is exploited via 
\begin{equation}
\label{analysis}
	{\bf{s^*}} = \underset{{\bf{s}} \in \mathbb{R}^{n}}{\arg \min} \quad g({\bf{\Omega s}}) \quad \text{s.t.} \quad \| {\bf{\Phi s - y}} \|_{2}^{2} \leq \epsilon.
\end{equation}
The analysis model has proven useful in the field of image reconstruction and we thus restrict ourselves to compressively sensed images here.
Our approach is motivated by the observation that learning the operator leads to an improved image reconstruction quality \cite{hawe-tip-12}, \cite{rubinstein-tsp-12} compared to applying a finite difference operator that approximates the image gradient, known as Total Variation (TV-norm) regularization \cite{candes-tit-06}, \cite{romberg-spm-08}. 
In contrast to the task of dictionary learning only a few analysis operator learning algorithms have been proposed in the literature so far, cf. \cite{roth-ijcv-2009}, \cite{yaghoobi-icassp-12}, \cite{rubinstein-tsp-12}, \cite{hawe-tip-12}. 
Furthermore, from image denoising it is known that the reconstruction accuracy can be further improved when the dictionary or operator is not only learned on some general and representative training set, but rather directly on the specific signal that has to be reconstructed \cite{elad-tip-06}, \cite{rubinstein-tsp-12}.
These observations prompted us to combine the image reconstruction performance of the analysis approach together with the accuracy improvement capabilities of a learned operator.

\subsection{Blind Compressive Sensing}
The principle of CS relies on the fact that the signal ${\bf{s}}$ has a sparse representation in a \emph{given} basis or dictionary ${\bf{\mathcal{D}}}$ that is universal for the considered signal class of interest. However, such universal dictionaries do not necessarily result in the sparsest possible representation, which is crucial for the recovery success. Due to this, in \cite{gleichman-tit-11} the concept of Blind Compressive Sensing (BCS) has been introduced, which aims at simultaneously learning the dictionary and reconstructing the signal, see also \cite{silva-arxiv-11} and \cite{studer-icassp-12} for an extension of this idea.
Note that all these methods are based on the synthesis model and consider the problem of finding a suitable dictionary, while in this paper we focus on the analysis model.

\subsection{Our Contribution}
\label{subsec:13}
In this work we address the problem of signal reconstruction from compressively sensed data regularized by an adaptively learned analysis operator.
The work of Hawe \textit{et al.} \cite{hawe-tip-12}, which focuses on learning a global patch based analysis operator from noise free training samples, has already shown the superior performance of a learned operator compared to 
state-of-the-art analysis and synthesis based regularization, like e.g. K-SVD denoising, in the context of classical image reconstruction problems. 
That is why we extend this idea and build on their work to utilize the learning process to obtain a signal dependent regularization of the inverse problem. 
Since we are dealing with compressive measurements, our approach can be interpreted as an analysis-based BCS problem with no prior knowledge about the operator.
 
We extend the algorithm proposed in \cite{hawe-tip-12}, where the operator is learned by a geometric Conjugate Gradient (CG) method on the so-called oblique manifold, to our setting of simultaneous image reconstruction and operator learning. This approach allows us to compensate for various sampling noise models, i.e. Gaussian or impulsive noise, by simply exchanging the data fidelity term. 
To summarize, the advantages of our approach are as follows:
 (i) The learning process allows to adaptively find an adequate operator that fits the underlying image structure.
 (ii) There is no necessity  to train the operator prior to the reconstruction.
 (iii) Different noise types are handled by simply exchanging the data fidelity term.

\section{Problem Statement}
\label{sec:2}
Our goal is to find a local analysis operator ${\bf{\Omega}} \in \mathbb{R}^{k \times n}$ with $k \geq n$ simultaneously to the signal ${\bf{s}} \in \mathbb{R}^{N}$ that has to be reconstructed from the compressive measurements.
Here, the vector ${\bf{s}}$ denotes a vectorized image of dimension $N = wh$, with $w$ being the width and $h$ being the height of the image, respectively, obtained by stacking the columns of the image above each other.
Note that the analysis operator has to be applied to local image patches rather than to the whole image. We denote the binary  $(n \times N)$ matrix that extracts the patch centered at the $(r,c)$ pixel by ${\bf{\mathcal{P}}}_{rc}$.
Furthermore, practice has shown that the learning process is significantly faster if \emph{centered}, i.e. zero mean patches are considered.
This can be easily incorporated by multiplying the vectorized patch with $\mathcal{M}:=(\mathcal{I}_{n \times n} -\frac{1}{n} \mathcal{J}_{n \times n})$, where $\mathcal{I}$ and $\mathcal{J}$ are the identity operator and the matrix with all elements equal to one, respectively.
We employ constant padding at the image borders, i.e. replicating the boundary pixel values.
In the end, we globally promote sparsity with an appropriate function $g( \cdot ): \mathbb{R}^{k} \to \mathbb{R}$ and write for the problem of finding a suitable analysis operator
\begin{equation}
\label{min_global_operator}
	{\bf{\Omega}}^{*} = \underset{{\bf{\Omega}} \in \mathcal{C}}{\arg \min} \sum_{r,c} g( {\bf{\Omega}} \, \mathcal{M}\,{\bf{\mathcal{P}}}_{rc} {\bf{s}} )^2 ,
\end{equation}
where $\mathcal{C}$ denotes an admissible set, which implies some constraints on ${\bf{\Omega}}$ to avoid trivial solutions. 
We follow the considerations of the authors in \cite{hawe-tip-12}, demanding that: 
\begin{compactenum}[(i)]
 \item  The rows of ${\bf{\Omega}}$ have unit Euclidean norm, i.e. $\| {\boldsymbol{\omega}}_{i} \|_2 = 1$, for $i = 1,...,k$, where ${\boldsymbol{\omega}}_{i}$ denotes the transposed of the $i^{th}$-row of ${\bf{\Omega}}$.
 \item The analysis operator ${\bf{\Omega}}$ has full rank, i.e. $\text{rk}({\bf{\Omega}}) = n$. 
 \item The mutual coherence of the analysis operator should be moderate.
\end{compactenum}
These constraints motivate to consider the set of full rank matrices with normalized columns, which admits a manifold structure known as the oblique manifold
\begin{equation}
\label{oblique_manifold}
	\text{OB}(n,k) := \{\mathcal{X} \in \mathbb{R}^{n \times k} \, | \, \text{rk}(\mathcal{X}) = n, \mathrm{ddiag}(\mathcal{X}^\top\mathcal{X}) = \mathcal{I}_{k}\}.
\end{equation} 
Here, $\mathrm{ddiag}({\bf{\mathcal{V}}})$ is the diagonal matrix whose entries on the diagonal are those of ${\bf{\mathcal{V}}}$.
Since we require the rows of ${\bf{\Omega}}$ to have unit Euclidean norm, we restrict ${\bf{\Omega}}^\top$ to be an element of $\text{OB}(n,k)$. 
To enforce the rank constraint (ii) we employ the penalty function
\begin{equation}
\label{logdetterm}
	h({\bf{\Omega}}) := - \tfrac{1}{n \log{(n)}} \log \det ( \tfrac{1}{k} {\bf{\Omega}}^\top {\bf{\Omega}} ).
\end{equation} 
Furthermore, the mutual coherence of the analysis operator, formulated in constraint (iii), can be controlled via the logarithmic barrier function of the atoms' scalar products, namely
\begin{equation}
\label{logbarrier}
	r({\bf{\Omega}}) := - \sum_{1 \leq i < j \leq k} \log ( 1 - ( {\boldsymbol{\omega}}_{i}^\top {\boldsymbol{\omega}}_{j} )^2 ).
\end{equation} 
Considerations concerning the usefulness of these penalty functions can be found in \cite{hawe-tip-12}.
To measure the sparsity of the analyzed patches, we use the differentiable sparsity promoting function
\begin{equation}
\label{sparsity_fct}
	g({\bf{w}}) = \sum_{j} \, \log \left( 1 + c \cdot ({\bf{e}}_{j}^{\top} {\bf{w}} )^2 \right) ,
\end{equation} 
where $c$ is a positive constant and ${\bf{e}}_{j}$ represents the $j^{th}$ standard basis vector with the same length as ${\bf{w}}$. 

Since we are interested in simultaneous operator learning and image reconstruction, we further introduce a data term $p( \cdot )$, which measures the fidelity of the reconstructed signal to the measurements ${\bf{y}} \in \mathbb{R}^{M}$.
The choice of $p(\cdot)$ depends on the error model, i.e. by using $p( \cdot ) = \| \cdot  \|_2^2$ the error is assumed to be Gaussian distributed.
If the noise is sparsely distributed over the measurements, we set $p( \cdot ) = g( \cdot )$. This error model has also been utilized in \cite{carillo-tisp-10} to compensate for sparse outliers in the measurements.

Finally, combining the data term with the constraints and the sparsity promoting function $g$, the augmented Lagrangian optimization problem for adaptively learning the analysis operator with simultaneous image reconstruction consists of minimizing the cost 
\begin{equation}
\label{opt_problem}
	 f({\bf{\Omega}}^{\top}\!,{\bf{s}})   = \quad \tfrac{1}{2 B} \sum_{(r,c)} \, g( {\bf{\Omega}} \, \mathcal{M}\, {\bf{\mathcal{P}}}_{rc} {\bf{s}} )^2  
	+  \eta \, p( {\bf{\Phi}} {\bf{s}} - {\bf{y}} )
  +  \gamma \, h({\bf{\Omega}}) 
	+  \kappa \, r({\bf{\Omega}}),
\end{equation} 
subject to ${\bf{\Omega}}^{\top} \in \text{OB}(n,k)$ with the measurement matrix ${\bf{\Phi}} \in \mathbb{R}^{M \times N}$. The scalar $B$ denotes the number of extracted image patches. 
The parameter $\eta \in \mathbb{R}^{+}$ weights the fidelity of the solution to the measurements and the parameters $\gamma, \kappa \in \mathbb{R}^{+}$ control the influence of the two constraints.

\section{Optimization Algorithm}
\label{sec:3}

Since the cost function \eqref{opt_problem} is restricted to a smooth manifold, we follow \cite{hawe-tip-12} and employ a conjugate gradient on manifolds approach to solve the optimization problem. 
The CG approach is scalable and converges fast in practice. It is thus well-suited to handle the high dimensional problem of simultaneous image reconstruction and operator learning.
The challenges for developing the CG method are the efficient computation of the Riemannian gradient, the step-size and the update directions. To that end, we employ the product manifold structure of $\text{OB}(n,k) \times \mathbb{R}^N$ considered as a Riemannian submanifold of $\mathbb{R}^{n \times k} \times \mathbb{R}^N$.
To enhance legibility in the remainder of this section we denote the oblique manifold by OB.
We further denote the tangent space at a point $\mathbf{\Omega}^\top=\mathcal{X} \in \text{OB}$ as $T_{\mathcal{X}} \text{OB}$, with $\Xi \in T_{\mathcal{X}} \text{OB}$ being a tangent vector at $\mathcal{X}$.

The Riemannian gradient at $\mathcal{X}$ is given by the orthogonal projection of the standard (Euclidean) gradient onto the tangent space $T_{\mathcal{X}} \text{OB}$.
The orthogonal projection of a matrix $\mathcal{Q} \in \mathbb{R}^{n \times k}$ onto the tangent space $T_{\mathcal{X}} \text{OB}$ is obtained by
$\Pi_{T_{\mathcal{X}} \text{OB}}(\mathcal{Q}) = \mathcal{Q} - \mathcal{X} \mathrm{ddiag}(\mathcal{X}^\top \mathcal{Q})$.
%
%
Using the product structure and denoting the partial derivatives of $f$ by $\nabla_{{\bf{s}}} f(\mathcal{X},{\bf{s}})$ and $\nabla_{\mathcal{X}} f({\mathcal{X}},{\bf{s}})$, respectively, the Riemannian gradient of the cost function is 
\begin{equation}
\label{riem_grad}
	 \mathcal{G}(\mathcal{X}, {\bf{s}} ) =\Big( \Pi_{T_{\mathcal{X}} \text{OB}}(\nabla_{\mathcal{X}} f), \nabla_{{\bf{s}}} f\Big).
\end{equation}

In CG methods the updated search directions $(\mathcal{H}^{(i+1)}, {\bf{h}}^{(i+1)}) \in T_{\mathcal{X}^{(i+1)}} \text{OB} \times \mathbb{R}^{N}$ are linear combinations of the respective gradient and the previous search directions $(\mathcal{H}^{(i)}, {\bf{h}}^{(i)}) \in T_{\mathcal{X}^{(i)}} \text{OB} \times \mathbb{R}^{N}$.
The identification of different tangent spaces is done by the so-called parallel transport  $\mathcal{T}_{\Xi}^{(i+1)} := \mathcal{T}(\Xi, \mathcal{X}^{(i)}, \mathcal{H}^{(i)}, \alpha^{(i)})$, which transports a tangent vector $\Xi$ along a \emph{geodesic} to the tangent space $T_{\mathcal{X}^{(i+1)}} \text{OB}$.
In the manifold setting geodesics can be considered as the generalization of straight lines. 
We denote the geodesic from $\mathcal{X}^{(i)}$ along the direction $\mathcal{H}^{(i)}$ as $\Gamma(\mathcal{X}^{(i)}, \mathcal{H}^{(i)}, t)$.
Regarding the product manifold the new iterates are computed by
\begin{equation}
	\label{newiterate}
  (\mathcal{X}^{(i+1)}, {\bf{s}}^{(i+1)}) = \left( \Gamma(\mathcal{X}^{(i)}, \mathcal{H}^{(i)}, \alpha^{(i)}), {\bf{s}}^{(i)} + \alpha^{(i)}{\bf{h}}^{(i)} \right),
\end{equation}
where $\alpha^{(i)}$ denotes the step size that leads to a sufficient decrease of the cost function.
The parallel transport along the geodesics in the product manifold is then given by 
\begin{equation}
  \label{paralleltransp}
  \mathcal{T}_{\mathcal{H}^{(i)}, {\bf{h}}^{(i)}}^{(i+1)} = \left( \mathcal{T}_{\mathcal{H}^{(i)}}^{(i+1)}, {\bf{h}}^{(i)} \right).
\end{equation}

We use a hybridization of the Hestenes-Stiefel (HS) and the Dai Yuan (DY) formula as motivated in \cite{dai-aor-01} to determine the update of the search direction.
With the shorthand notations $\mathcal{G}^{(i)} := \mathcal{G}(\mathcal{X}^{(i)})$ and ${\bf{g}}^{(i)} := \mathcal{G}({\bf{s}}^{(i)})$, as well as $\mathcal{U}^{(i+1)} = \mathcal{G}^{(i+1)} - \mathcal{T}_{\mathcal{G}^{(i)}}^{(i+1)}$ and ${\bf{u}}^{(i+1)} = {\bf{g}}^{(i+1)} - {\bf{g}}^{(i)}$ the manifold adaptions of these formulas are
\begin{align}
	\label{cgupdate1}
  \beta_{\text{HS}}^{(i)} &= \frac{\langle \mathcal{G}^{(i+1)}, \mathcal{U}^{(i+1)} \rangle + \langle {\bf{g}}^{(i+1)}, {\bf{u}}^{(i+1)} \rangle}{\langle \mathcal{T}_{\mathcal{H}^{(i)}}^{(i+1)}, \mathcal{U}^{(i+1)} \rangle + \langle {\bf{h}}^{(i)}, {\bf{u}}^{(i+1)} \rangle}, \\
	\label{cgupdate2}
  \beta_{\text{DY}}^{(i)} &= \frac{\langle \mathcal{G}^{(i+1)}, \mathcal{G}^{(i+1)} \rangle + \langle {\bf{g}}^{(i+1)}, {\bf{g}}^{(i+1)} \rangle}{\langle \mathcal{T}_{\mathcal{H}^{(i)}}^{(i+1)}, \mathcal{U}^{(i+1)} \rangle + \langle {\bf{h}}^{(i)}, {\bf{u}}^{(i+1)} \rangle},
\end{align}
where $\langle \cdot , \cdot \rangle$ denotes the standard inner product in the respective Euclidean spaces.
With the hybrid update formula 
\begin{equation}
	\label{hybupdate}
	\beta_{\text{hyb}}^{(i)}=\max \left( 0, \min(\beta_{\text{DY}}^{(i)},\beta_{\text{HS}}^{(i)}) \right),
\end{equation}	
the new search directions are given by
\begin{equation}
  \label{searchdirections}
  (\mathcal{H}^{(i+1)}, {\bf{h}}^{(i+1)}) = \left( -\mathcal{G}(\mathcal{X}^{(i+1)},{\bf{s}}^{(i+1)}) + \beta_{\text{hyb}}^{(i)} \mathcal{T}_{\mathcal{H}^{(i)}, {\bf{h}}^{(i)}}^{(i+1)} \right).
\end{equation}
In our implementation we use the well-known backtracking line search which is adapted to the manifold setting until the Armijo condition is met.
We name our method Analysis Blind Compressive Sensing (ABCS) and briefly summarize the whole procedure in Algorithm 1.
For further details concerning CG-methods on the oblique manifold the reader is referred to \cite{hawe-tip-12}, \cite{absil-puc-08}, and \cite{kleinsteuber-spl-12}. 

\noindent
\mbox{}\hrulefill

\vspace{-.3em}
\noindent
\textbf{Algorithm 1} ABCS 

\vspace{-.8em}
\noindent
\mbox{}\hrulefill

\noindent
\textbf{Input:} Initial operator ${\bf{\Omega}}_{init}$, noisy measurements ${\bf{y}}$, measurement matrix ${\bf{\Phi}}$, parameters $\gamma,\kappa,\eta,c$ \\
\textbf{Set:} $i \leftarrow 0$, ${\bf{s}}^{(0)} \leftarrow {\bf{\Phi}}^{\top} {\bf{y}}$, $\mathcal{X}^{(0)} \leftarrow {\bf{\Omega}}_{init}^{\top}$, $\mathcal{H}^{(0)} \leftarrow -\mathcal{G}^{(0)}$, ${\bf{h}}^{(0)} \leftarrow -{\bf{g}}^{(0)}$ 
\begin{algorithmic}[1]
	\Repeat
	\State perform backtracking line search to get step size $\alpha^{(i)}$
	\State update to $(\mathcal{X}^{(i+1)}, {\bf{s}}^{(i+1)} )$, cf. \eqref{newiterate}
	\State compute $\mathcal{G}(\mathcal{X}^{(i+1)}, {\bf{s}}^{(i+1)} )$
	\State compute $\beta^{(i)}_{\text{hyb}}$, cf. \eqref{hybupdate} 
	\State compute new CG-search directions $(\mathcal{H}^{(i+1)}, {\bf{h}}^{(i+1)})$	
	\State $i = i + 1$
	\Until $\| \mathcal{X}^{(i)} - \mathcal{X}^{(i-1)} \|_{F} < 10^{-6} \vee i = $ maximum \# of iterations
\end{algorithmic}
\textbf{Output:} ${\bf{\Omega}}^{*} \leftarrow \mathcal{X}^{(i) \top}$, ${\bf{s}}^{*} \leftarrow {\bf{s}}^{(i)}$

\vspace{-.8em}
\noindent
\mbox{}\hrulefill

\section{Experimental Results}
\label{sec:4}

\begin{figure}[t]
\centering
\subfigure[ABCS]{\includegraphics[width=0.45\columnwidth]{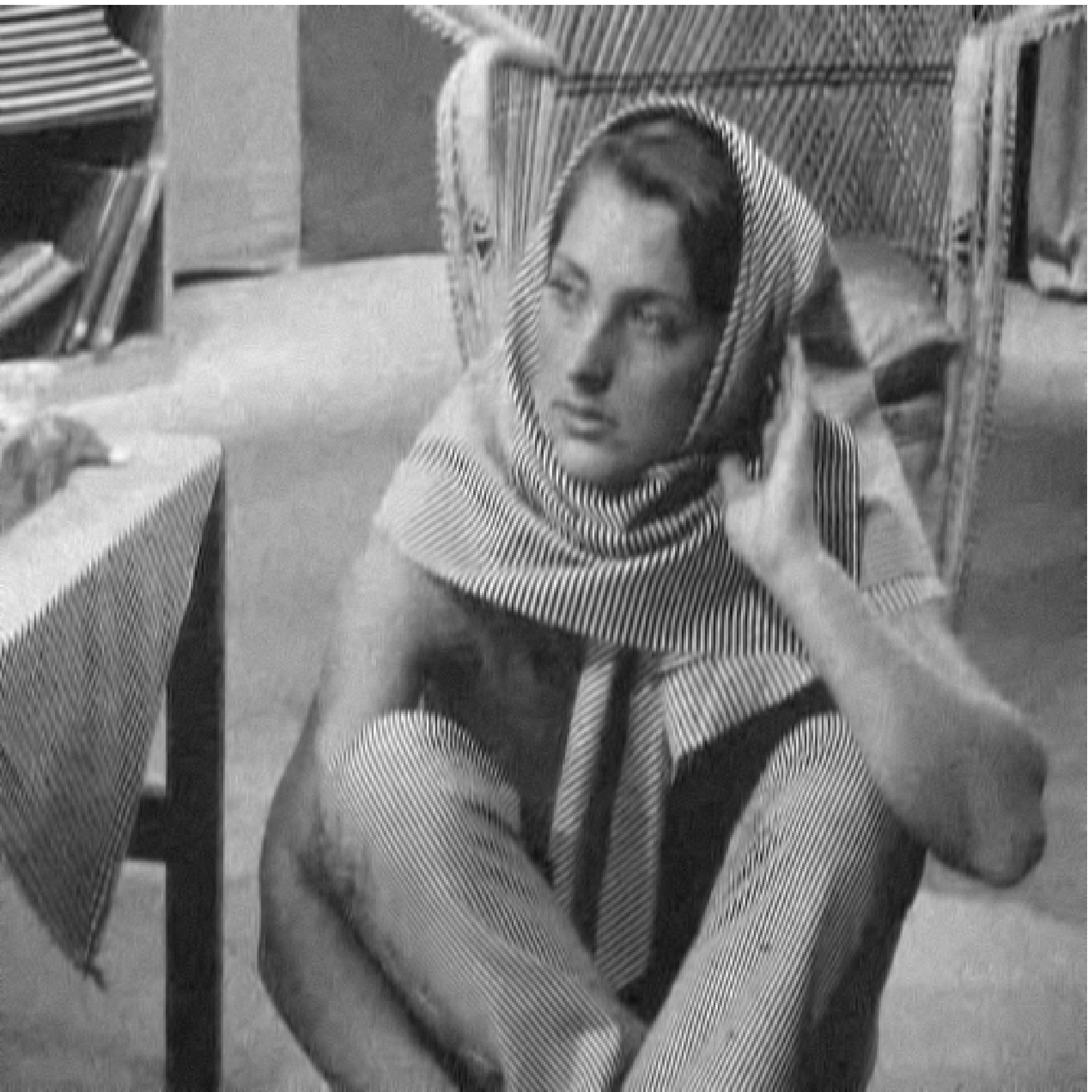}}\quad
\subfigure[NESTA+TV]{\includegraphics[width=0.45\columnwidth]{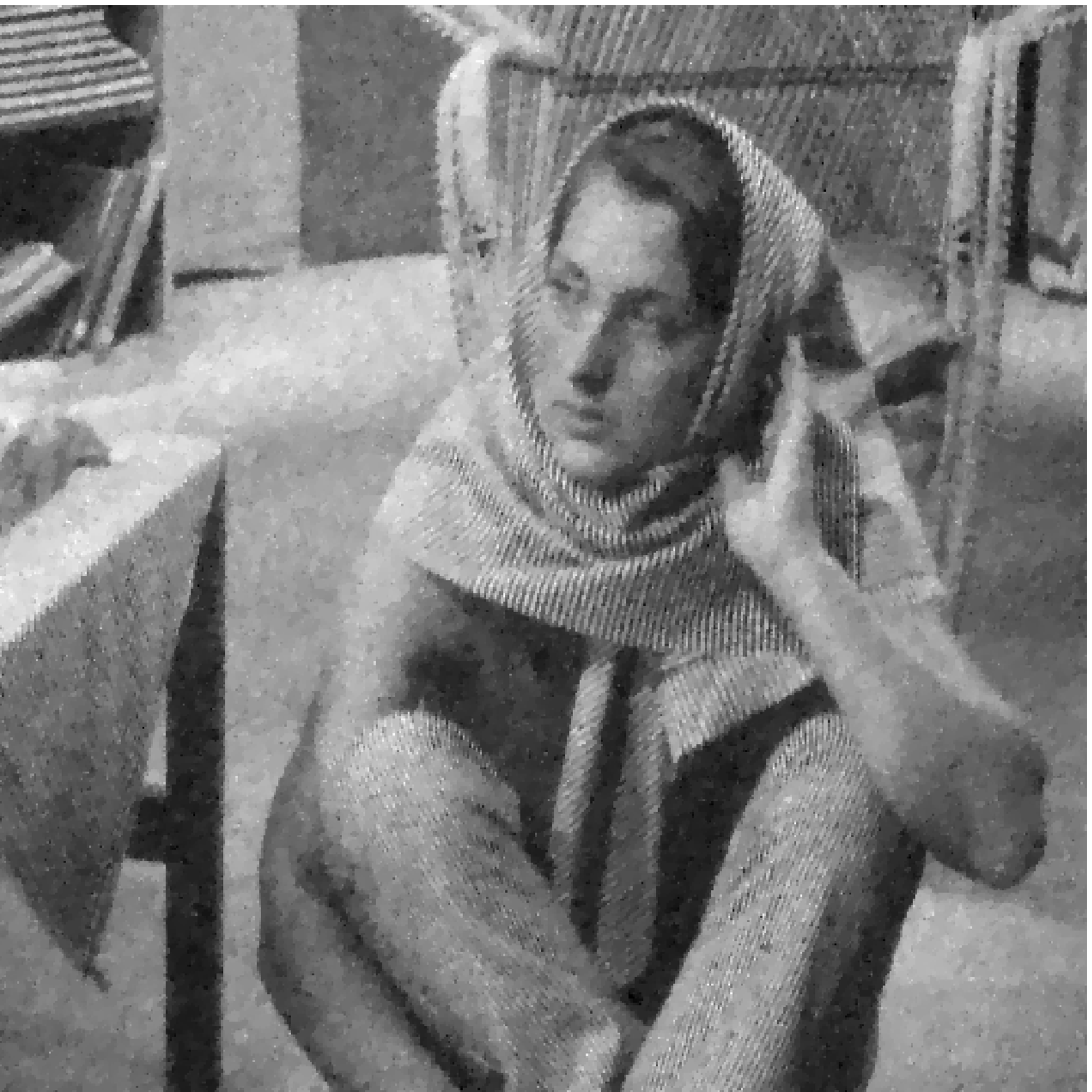}}
\caption{Reconstruction of the \textit{Barbara} image from $M = N/4$ measurements corrupted by additive white Gaussian noise with $\sigma_{\text{noise}} = 5.1$.}
\label{fig:res}
\end{figure}

To measure the image reconstruction accuracy we use the peak signal-to-noise ratio \linebreak ${(\textit{PSNR}) =  10 \log ( 255^2 N / \sum_{i = 1}^{N} (s_{i} -s_{i}^{*})^2)}$ and the Mean Structural SIMilarity Index (\textit{MSSIM}), with the same set of parameters as originally suggested and implemented in \cite{wang-tip-04}.
Throughout our experiments we use a patch size of $(7 \times 7)$, i.e. $n = 49$ and set $k = 2n$, as larger values of $k$ do not enhance the reconstruction quality.
We initialized ${\bf{\Omega}}_{init}$ to be a random matrix and normalized the rows to unit norm. With this initialization, convergence to a local minimum was observed in all our experiments.
The parameters for the constraints are set to $\gamma = 20$ and $\kappa = 1000$. 
The constant $c$ in the sparsity inducing function \eqref{sparsity_fct} is chosen as $c = 10^4$. 
The parameter $\eta$ takes into account the size of the image as well as the operator size and reads $\eta = \hat \eta \cdot \left( \tfrac{k}{L n} \right)^2$, with $\hat \eta$ adjusted according to the noise level as explained below and a normalization factor $L = \tfrac{\sqrt{N}}{256}$.

We evaluate our method on the three images \textit{Girl} ($256 \times 256$), \textit{Barbara} ($512 \times 512$), and \textit{Texture}\footnote{Image 1.5.03.tiff obtained from the USC-SIPI Image Database: http://sipi.usc.edu/database/ and cropped to ($256 \times 256$)} ($256 \times 256$).
The measurements are obtained by using the real valued noiselet transformation proposed in \cite{romberg-spm-08}.

In the first experiment we show the robustness of the ABCS algorithm to sampling noise which follows a Gaussian distribution.
For this purpose, the measurements have been artificially corrupted by additive white Gaussian noise with standard deviation $\sigma_{\text{noise}}$.
The data term in \eqref{opt_problem} reads $p( \cdot ) = \| \cdot \|^{2}_{2}$.
We assume the noise level $\sigma_{\text{noise}}$ to be known and set $\hat \eta = \tfrac{1000}{\sigma_{\text{noise}}}$.
Two measurement rates $M = N/4$ and $M = N/10$ are considered.
Table \ref{tab:denoising} shows the reconstruction performance for different noise levels. 
For comparison we used the algorithm of \cite{becker-siam-11} (NESTA), with TV-norm regularization and optimized parameters.
Figure \ref{fig:res} shows the reconstructed images from $M = N/4$ measurements and a noise level of $\sigma_{\text{noise}} = 5.1$. 
We also tested the algorithm proposed in \cite{li-cam-12} (TVAL3) with different parameters, which achieves results comparable to NESTA. Due to space limitations, detailed results are not listed here. 
In all settings, the same measurements are used.
\begin{table}[t]
\footnotesize
\renewcommand{\arraystretch}{1.3}
\caption{Image reconstruction from measurements corrupted by additive white Gaussian noise with standard deviation $\sigma_{\text{noise}}$. The measurement rates are $M = N/4$ (top) and $M = N/10$ (bottom). Achieved PSNR in decibels and MSSIM.}
\label{tab:denoising}
\centering
\begin{tabular}{c|c||c|c||c|c||c|c}
\hline
 & & \multicolumn{2}{c||}{\textit{Girl}} & \multicolumn{2}{c||}{\textit{Barbara}} & \multicolumn{2}{c}{\textit{Texture}}  \\
\cline{3-8}
Method    &  $\sigma_{\text{noise}}$ & \textit{PSNR} & \textit{MSSIM} & \textit{PSNR} &  \textit{MSSIM}  & \textit{PSNR} &  \textit{MSSIM}   \\
\hline\hline
NESTA     &       0.1                &   31.97       &   0.794        &   25.03       &   0.686          &   26.92       &   0.732           \\
+TV       &       5.1                &   30.97       &   0.754        &   24.71       &   0.676          &   26.53       &   0.717           \\
          &       10.2               &   29.72       &   0.701        &   24.01       &   0.641          &   25.66       &   0.668           \\ \hline
ABCS      &       0.1                &   32.38       &   0.806        &   32.10       &   0.895          &   28.33       &   0.807           \\
          &       5.1                &   31.36       &   0.767        &   29.79       &   0.847          &   27.73       &   0.779           \\
          &       10.2               &   29.94       &   0.708        &   27.39       &   0.766          &   26.59       &   0.731           \\ \hline \hline
NESTA     &       0.1                &   29.51       &   0.690        &   22.59       &   0.560          &   23.87       &   0.544           \\
+TV       &       5.1                &   28.95       &   0.667        &   22.56       &   0.576          &   23.72       &   0.539           \\
          &       10.2               &   28.06       &   0.632        &   22.31       &   0.559          &   23.32       &   0.522           \\ \hline
ABCS      &       0.1                &   29.98       &   0.711        &   24.60       &   0.651          &   25.06       &   0.644           \\
          &       5.1                &   29.29       &   0.684        &   23.31       &   0.587          &   24.84       &   0.625           \\
          &       10.2               &   28.31       &   0.645        &   22.79       &   0.557          &   24.15       &   0.590           \\ \hline					
\hline
\end{tabular}
\end{table}

To handle measurements that are corrupted by impulsive noise, we exchange the data fidelity function in \eqref{opt_problem} to $p( \cdot ) = g( \cdot )$ in our second experiment.
Corrupted coefficients are set to a value of $\pm \, 1.25 \cdot \vert y \vert_{max}$.
Table \ref{tab:spdenoising} summarizes the results for a sampling rate of $M = N/4$ and different amounts $d$ of corrupted measurements.
In the ABCS algorithm, the parameter $\hat \eta$ is set to $0.08$ and to $0.05$ for $10\%$ and, respectively $20\%$, of corrupted measurements. 
To compare our results achieved with the adaptively learned operator, we used the same setting of the reconstruction scheme with a fixed finite difference operator denoted as TV in Table \ref{tab:spdenoising}. 
\begin{table}[!t]
\footnotesize
\renewcommand{\arraystretch}{1.3}
\caption{Image reconstruction from measurements corrupted by impulsive noise. Achieved PSNR in decibels and MSSIM. The values in brackets correspond to the amount of corrupted measurements.}
\label{tab:spdenoising}
\centering
\begin{tabular}{c||c|c||c|c||c|c}
\hline
  & \multicolumn{2}{c||}{\textit{Girl}} & \multicolumn{2}{c||}{\textit{Barbara}} & \multicolumn{2}{c}{\textit{Texture}}  \\
\cline{2-7}
Method           & \textit{PSNR} & \textit{MSSIM} & \textit{PSNR} &  \textit{MSSIM}  & \textit{PSNR} &  \textit{MSSIM}   \\
\hline\hline
ABCS (10\%)       &   31.82       &   0.784        &   28.78       &    0.827         &   26.66       &   0.719           \\
ABCS (20\%)       &   30.89       &   0.749        &   22.47       &    0.577         &   25.18       &   0.617           \\
TV (10\%)         &   30.23       &   0.727        &   22.80       &    0.532         &   24.13       &   0.585           \\
TV (20\%)         &   29.80       &   0.708        &   22.43       &    0.510         &   23.35       &   0.537           \\
\hline
\end{tabular}
\end{table}

Both experiments confirm that the adaptively learned operator leads to an accuracy improvement compared to the reconstruction quality obtained with a fixed finite difference operator. 
In particular, the structures in the \textit{Barbara} and \textit{Texture} image are better preserved by ABCS.

\section{Conclusion}
\label{sec:5}
In this article we proposed an analysis based blind compressive sensing algorithm that simultaneously reconstructs an image from compressively sensed data and learns an appropriate analysis operator. This process is formulated as an optimization problem, which is tackled via a geometric conjugate gradient approach that updates both the operator and the image as a whole at each iteration. Furthermore, the algorithm can be easily adapted to different noise models by simply exchanging the data fidelity term.

%



\footnotesize{
\bibliographystyle{ieeetr}
\bibliography{ABCSbibfile}}

\begin{thebibliography}{10}

\bibitem{elad-invprob-07}
M.~Elad, P.~Milanfar, and R.~Rubinstein, ``Analysis versus synthesis in signal
  priors,'' {\em Inverse Problems}, vol.~23, no.~3, pp.~947--968, 2007.

\bibitem{tropp-poti-2010}
J.~A. Tropp and S.~J. Wright, ``Computational methods for sparse solution of
  linear inverse problems,'' {\em Proceedings of the IEEE}, vol.~98, no.~6,
  pp.~948--958, 2010.

\bibitem{hawe-tip-12}
S.~Hawe, M.~Kleinsteuber, and K.~Diepold, ``Analysis operator learning and its
  application to image reconstruction,'' {\em IEEE Transactions on Image
  Processing}, 2013.
\newblock published online.

\bibitem{rubinstein-tsp-12}
R.~Rubinstein, T.~Peleg, and M.~Elad, ``Analysis k-svd: A dictionary-learning
  algorithm for the analysis sparse model,'' {\em IEEE Transactions on Signal
  Processing}, vol.~61, no.~3, pp.~661--677, 2013.

\bibitem{candes-tit-06}
E.~J. Cand{\`e}s, J.~Romberg, and T.~Tao, ``Robust uncertainty principles:
  exact signal reconstruction from highly incomplete frequency information,''
  {\em IEEE Transactions on Information Theory}, vol.~52, no.~2, pp.~489--509,
  2006.

\bibitem{romberg-spm-08}
J.~Romberg, ``Imaging via compressive sampling,'' {\em IEEE Signal Processing
  Magazine}, vol.~25, no.~2, pp.~14--20, 2008.

\bibitem{roth-ijcv-2009}
S.~Roth and M.~Black, ``Fields of experts,'' {\em International Journal of
  Computer Vision}, vol.~82, no.~2, pp.~205--229, 2009.

\bibitem{yaghoobi-icassp-12}
M.~Yaghoobi, S.~Nam, R.~Gribonval, and M.~E. Davies, ``Noise aware analysis
  operator learning for approximately cosparse signals,'' in {\em IEEE
  International Conference on Acoustics, Speech and Signal Processing
  (ICASSP)}, pp.~5409--5412, 2012.

\bibitem{elad-tip-06}
M.~Elad and M.~Aharon, ``Image denoising via sparse and redundant
  representations over learned dictionaries,'' {\em IEEE Transactions on Image
  Processing}, vol.~15, no.~12, pp.~3736--3745, 2006.

\bibitem{gleichman-tit-11}
S.~Gleichman and Y.~C. Eldar, ``Blind compressed sensing,'' {\em IEEE
  Transactions on Information Theory}, vol.~57, no.~10, pp.~6958--6975, 2011.

\bibitem{silva-arxiv-11}
J.~Silva, M.~Chen, Y.~C. Eldar, G.~Sapiro, and L.~Carin, ``Blind compressed
  sensing over a structured union of subspaces.'' arXiv:1103.2469v1, 2011.

\bibitem{studer-icassp-12}
C.~Studer and R.~Baraniuk, ``Dictionary learning from sparsely corrupted or
  compressed signals,'' in {\em IEEE International Conference on Acoustics,
  Speech and Signal Processing (ICASSP)}, pp.~3341--3344, 2012.

\bibitem{carillo-tisp-10}
R.~E. Carrillo, K.~E. Barner, and T.~C. Aysal, ``Robust sampling and
  reconstruction methods for sparse signals in the presence of impulsive
  noise,'' {\em IEEE Journal of Selected Topics in Signal Processing}, vol.~4,
  no.~2, pp.~392--408, 2010.

\bibitem{dai-aor-01}
Y.~Dai and Y.~Yuan, ``An efficient hybrid conjugate gradient method for
  unconstrained optimization,'' {\em Annals of Operations Research}, vol.~103,
  no.~1-4, pp.~33--47, 2001.

\bibitem{absil-puc-08}
P.-A. Absil, R.~Mahony, and R.~Sepulchre, {\em Optimization Algorithms on
  Matrix Manifolds}.
\newblock Princeton, NJ: Princeton University Press, 2008.

\bibitem{kleinsteuber-spl-12}
M.~Kleinsteuber and H.~Shen, ``Blind source separation with compressively
  sensed linear mixtures,'' {\em IEEE Signal Processing Letters}, vol.~19,
  no.~2, pp.~107--110, 2012.

\bibitem{wang-tip-04}
Z.~Wang, A.~C. Bovik, H.~R. Sheikh, and E.~P. Simoncelli, ``Image quality
  assessment: from error visibility to structural similarity,'' {\em IEEE
  Transactions on Image Processing}, vol.~13, no.~4, pp.~600--612, 2004.

\bibitem{becker-siam-11}
S.~Becker, J.~Bobin, and E.~J. Cand{\`e}s, ``Nesta: A fast and accurate
  first-order method for sparse recovery,'' {\em SIAM Journal on Imaging
  Sciences}, vol.~4, no.~1, pp.~1--39, 2011.

\bibitem{li-cam-12}
C.~Li, W.~Yin, H.~Jiang, and Y.~Zhang, ``An efficient augmented lagrangian
  method with applications to total variation minimization,'' Tech. Rep.
  TR12-13, Computational and Applied Mathematics, Rice University, Houston, TX,
  2012.

\end{thebibliography}

\end{document}